\documentclass[10pt]{article}

\begin{document}
\noindent
{\large UW-PT-98-05}\hfill{\large hep-lat/9801005}
\vspace{3pt}

\begin{center}
\LARGE{\bf Proposal for\\Topologically Unquenched QCD}
\end{center}
\vspace{3pt}

\begin{center}
\large{\bf Stephan D\"{u}rr}\\
\vspace{3pt}
{\sl Physics Department, University of Washington}\\
{\sl Box 351560, Seattle WA 98195, U.S.A.}\\
\large{\tt durr@phys.washington.edu}
\end{center}
\vspace{3pt}

\begin{abstract}
\noindent
A proposal is presented for simulating an improvement on quenched QCD with
dynamical fermions which interact with the gluon configuration only via the
topological index of the latter. Strengths and shortcomings of the method are
discussed and it is argued that the approximation --~though being crude~--
shares some qualitative aspects of full QCD which relate to the issue of chiral
symmetry breaking.
\end{abstract}


\newcommand{\pad}{\partial}
\newcommand{\pas}{\partial\!\!\!/}
\newcommand{\Dsl}{D\!\!\!\!/\,}
\newcommand{\Psl}{P\!\!\!\!/\;\!}
\newcommand{\hqu}{\hbar}
\newcommand{\ovr}{\over} 
\newcommand{\hal}{{1\ovr2}}
\newcommand{\til}{\tilde}
\newcommand{\pri}{^\prime}
\renewcommand{\dag}{^\dagger}
\newcommand{\<}{\langle}
\renewcommand{\>}{\rangle}
\newcommand{\lan}{\langle}
\newcommand{\gaf}{\gamma_5}
\newcommand{\lap}{\triangle}
\newcommand{\trc}{\rm tr}
\newcommand{\al}{\alpha}
\newcommand{\be}{\beta}
\newcommand{\ga}{\gamma}
\newcommand{\de}{\delta}
\newcommand{\ep}{\epsilon}
\newcommand{\ve}{\varepsilon}
\newcommand{\ze}{\zeta}
\newcommand{\et}{\eta}
\renewcommand{\th}{\theta}
\newcommand{\vt}{\vartheta}
\newcommand{\io}{\iota}
\newcommand{\ka}{\kappa}
\newcommand{\la}{\lambda}
\newcommand{\rh}{\rho}
\newcommand{\vr}{\varrho}
\newcommand{\si}{\sigma}
\newcommand{\ta}{\tau}
\newcommand{\ph}{\phi}
\newcommand{\vp}{\varphi}
\newcommand{\ch}{\chi}
\newcommand{\ps}{\psi}
\newcommand{\om}{\omega}
\newcommand{\psb}{\overline{\psi}}
\newcommand{\etb}{\overline{\eta}}
\newcommand{\psd}{\psi^{\dagger}}
\newcommand{\etd}{\eta^{\dagger}}
\newcommand{\rch}{{\rm ch}}
\newcommand{\rsh}{{\rm sh}}
\newcommand{\beq}{\begin{equation}}
\newcommand{\eeq}{\end{equation}}
\newcommand{\bdm}{\begin{displaymath}}
\newcommand{\edm}{\end{displaymath}}
\newcommand{\bea}{\begin{eqnarray}}
\newcommand{\eea}{\end{eqnarray}}


\section{Motivation}

Lattice Field Theory is the only viable technical scheme which allows one to
calculate low energy observables in QCD from first principles.
For phenomenological applications the lattice spacing $a$ has to be a fraction
of a hadron radius and the physical box-length $L$ should be large compared to
the Compton-wavelength of the low-energy (Goldstone-) modes.
Full QCD calculations to be run at phenomenological quark-mass values
exceed the present computers abilities.

The most serious problem stems from the fermion functional determinant, the
logarithm of which is a nonlocal contribution to the gluonic effective action.
This nonlocality --~which is more severe the lighter the quark-mass~-- tends to
slow down current algorithms dramatically.

In order to reduce the number of degrees of freedom to be handled numerically,
many computations in Lattice QCD have been performed in the quenched
approximation, where this determinant is replaced by one \cite{Hamber}, or,
more recently, in the partially quenched approximation, where the dynamical
(sea-) quarks are given a higher mass than the external (valence-) quarks
\cite{SESAM}.
Thus (partial) quenching amounts to suppressing the contribution of all
internal fermion loops in QCD by giving the quarks unphenomenologically high
or infinite masses.

Attempts to introduce the corresponding modifications in the low-energy
theory artificially in order to learn how to correct for them --~the results
being ``quenched'' and ``partially quenched Chiral Perturbation Theory''~--
have shown that the (partially) quenched approximation is, in some aspects,
fundamentally different from the full theory:
Quenched (euclidean) QCD was found not to have a Minkowski-space counterpart
\cite{Morel} and --~at least in principle~-- numerical results have to
be corrected for the occurrency of ``enhanced chiral logarithms'' 
\cite{QuenchedQCD, PartiallyQuenchedQCD}.
In addition, the $\et\pri$ was found to be a pseudo-Goldstone boson in
quenched QCD (as opposed to the situation in QCD) and its propagator shows
--~in case the low-energy analysis is correct~-- a double-pole right at the
same position in the $p^2$-plane as its single-pole \cite{QuenchedQCD}.
In a field-theoretic framework single poles and cuts in Green's functions are
associated with particles and their interactions, but there is no way a
multiple pole could be identified with a particle. 

Besides these somewhat theoretical challenges, there is a very practical
problem encountered in quenched simulations:
When a correlation-function of opera\-tors involving fermionic fields is
measured on a set of quenched configurations the entire measurement is
dominated (in particular in the limit of small quark-masses and/or strong
gauge-coupling) by a single or a limited number of configurations.
These so-called ``exceptional configurations'' have to be removed from the
sample and there is a strong theoretical motivation for doing so: Their
``exceptional'' behaviour results from spurious almost-zero real modes of the
Dirac operator which in full QCD would give rise to almost-zero factors in the
determinant and thus to a strong suppression of these configurations \cite{MQA}.
Hence removing the ``exceptional configurations'' is the right thing to do, but
the problem is that there is no canonical definition of how much excess is
required to make a configuration an ``exceptional'' one (and obviously this
choice influences the result of the measurement).

In the following, we shall propose an alternative to quenched QCD which amounts
to including a part of the functional determinant which can be evaluated with
limited computational costs.
From analytical considerations we will argue that it is not unreasonable to
hope that this approximation --~though being crude~-- gets some basic features
of full QCD qualitatively right: The ``topological'' part of the functional
determinant is sufficient to get symmetry restoration when the chiral limit is
performed in a finite volume and configurations get suppressed by a standard
determinant which accounts for nothing but the number of (lattice-descendents
of) zero-modes of the Dirac operator on that configuration -- which is less
than perfect but better than no suppression at all.


\section{Topologically Unquenched QCD}

We start from the generating functional for (euclidean) QCD with quark-masses
$m_i$, vacuum angle $\th$ and external fermionic currents $\et,\etb$
\beq
Z_\th[\etb\!,\!\et]\!=\!N\!\cdot\!\!\int\!DAD\psb\!D\ps\
e^{-\{\int\!{1\ovr4}GG+\int\!\psb(\Dsl+M)\ps-i\th\!\int\!{g^2\ovr32\pi^2}
G\til G+\int\!\psb\et+\int\!\etb\ps\}}
\label{one}
\eeq
where $\Dsl=\ga_\mu(\pad_\mu-igA_\mu)$ is the (euclidean) Dirac operator,
$\til G_{\mu\nu}={1\ovr2}\ep_{\mu\nu\si\rh} G_{\si\rh}$ is the dual of the
field-strength operator and the measure $DA$ is meant to include gauge fixing
and Faddeev Popov terms.
Here and in subsequent formulae we shall adopt the convention that the
quark-mass matrix $M$ is diagonal, positive-real and of rank $N_{\!f}$ (i.e.
the CP-violation stemming entirely from $\th$ if $\th\neq\pi{\rm{\bf Z}}$) and
the normalization factor $N$ is such as to guarantee $Z_0[0,0]=1$.
Integrating over the fermionic degrees of freedom one gets the usual formula
\beq
Z_\th[\etb,\et]=N\!\cdot\!\!
\int\!DA\ {\det(\Dsl\!+\!M)\ovr\det(\,\pas\!+\!M)}\
e^{\etb\!(\Dsl+M)^{-1}\et}\
e^{-\!\int\!{1\ovr4}GG\;+i\th\!\int\!{g^2\ovr32\pi^2}G\til G}
\label{two}
\eeq
where $\det(\!\Dsl\!+\!M)\!=\!\prod_{i=1}^{N_f}\!\det(\!\Dsl\!+\!m_i),\ \etb
(\Dsl\!+\!M)^{-1}\et\!=\!\sum_{i=1}^{N_f}\et_{\!(i)}\dag\!(\Dsl\!+\!m_i)^{-1}
\et_{\!(i)}$ and a factor which does not depend on the gauge field to be
integrated over has been pulled out of the normalizing constant.

As is known, the $SU(2)$-gauge field configurations on ${\bf R}^4$ with finite
action boundary condition or on the torus ${\bf T}^4$ fall into inequivalent
topological classes (labeled by an index $\nu\!\in\!{\bf Z}$) and a theorem by
Bott guarantees that this holds true for any simple Lie group, in particular
for all $SU(N_c)$ theories \cite{Topology}.
In a given sector $\nu$, any gauge field $A^{(\nu)}$ has an action which is
bounded from below by $8\pi^2/g^2\cdot\vert\nu\vert$.
Choosing an arbitrary ``standard configuration'' $A^{(\nu)}_{\rm std}$ in each
sector the generating functional (\ref{two}) can be rewritten as
\beq
Z_\th[\etb,\et]=N\cdot\!
\sum_{\nu\in{\rm\bf Z}}
{\det(\Dsl^{(\nu)}_{\rm std}\!\!+\!\!M)\ovr\det(\,\pas+\!M)}\,
e^{i\nu\th}\cdot\!
\int\!\!D\!A^{(\nu)}\,
{\det(\Dsl^{(\nu)}\!\!+\!\!M)\ovr\det(\Dsl^{(\nu)}_{\rm std}\!\!+\!\!M)}\,
e^{\etb(\Dsl^{(\nu)}\!+M)^{-1}\et}\,
e^{-\!\!\int\!\!{1\ovr4}GG}\!
\label{thr}
\eeq
where $\Dsl^{(\nu)}$ refers to the actual $A^{(\nu)}$ whereas
$\Dsl^{(\nu)}_{\rm std}$ refers to $A^{(\nu)}_{\rm std}$.

Equation (\ref{thr}) explicitly splits the full determinant in
(\ref{two}) into two (individu\-ally gauge-invariant) factors:
The second factor $\det((\Dsl^{(\nu)}\!+\!M)/(\Dsl^{(\nu)}_{\rm std}\!+\!M))$
--~which remains inside the integral~-- describes the deviation of the actual
gauge field $A^{(\nu)}$ from the standard configuration $A^{(\nu)}_{\rm std}$
into which it may be continuously deformed.
The first factor $\det((\Dsl^{(\nu)}_{\rm std}\!+\!M)/(\,\pas\!+\!M))$ --~which
depends on $\nu$ only and thus may be pulled out of the integral~-- accounts
for the topology of $A^{(\nu)}$.
Equation (\ref{thr}) makes apparent that quenched QCD actually does two
modifications simultaneously: It sets both determinant factors equal one.
In contrast, partially quenched QCD keeps both of them at the price of using
unphenomenologically high quark-masses.

An interesting alternative could be to keep the first (``topological'') factor
--~with exactly the same quark-mass as in the propagator sandwiched between the
currents $\etb,\et$~-- and to replace the second (``continuous'') factor by one.
The result defines ``topologically unquenched QCD'':
\beq
Z_{\th,\{A^{(\nu)}_{\rm std}\}}^{\rm TU}[\etb,\et]=N\cdot\!
\sum_{\nu\in{\rm\bf Z}}
{\det(\Dsl^{(\nu)}_{\rm std}\!\!+\!\!M)\ovr\det(\,\pas\!\!+\!\!M)}\
e^{i\nu\th}\cdot\!
\int\!\!D\!A^{(\nu)}\;e^{\etb(\Dsl^{(\nu)}\!+M)^{-1}\et}\;
e^{-\!\!\int\!\!{1\ovr4}GG}\ .
\label{for}
\eeq
Initially, the motivation to treat the two determinants in (\ref{thr}) on
unequal footing is an economic one: The ``topological'' factor in (\ref{thr})
is universal for all configurations within one class.
Moreover it bears the knowledge about the nontrivial topological structure of
QCD.
On the other hand, the ``continuous'' factor in (\ref{thr}) causes a dramatic
slowdown in numerical computations, since this determinant (or its change)
has to be computed for each configuration individually.

Note that the full-QCD generating functional (\ref{thr}) is independent of
the reference-configurations, whereas the ``topologically unquenched''
truncation (\ref{for}) does depend on this choice.
Later, we shall get rid of this ambiguity by describing two alternative
strategies for choosing the background to be used as a reference-configuration
in a given topological sector.

Note further that there is, a priori, no evidence that the
approximation~(\ref{for}) should be particularly good, i.e. it is not obvious
that the ``topological'' determinant (even with the best possible choice for
the reference-backgrounds) should be close to the full determinant.
Nevertheless, it is possible to show that the present analytical knowledge
about QCD in a finite box indicates that it is at least not illegitimate to
hope that including the ``topological'' part of the determinant may result in
a approximation to full QCD which is worth studying.


\section{Implemention with Stepwise Refinement}

Suppose we are provided a set of quenched configurations and we want to measure
a correlation function built from fermionic external currents.
Then the procedure according to ``topologically unquenched QCD'' would be as
follows:
\begin{enumerate}
\item
Classify the configurations w.r.t. their topological indices $\nu$
(using a method you consider both trustworthy and efficient).
\item
Use the gauge action you trust to compute, in each class, the gauge-action of
every configuration as well as the class-average $\bar S^{(\nu)}$ and choose the
reference-configuration according to one of the following two prescriptions:
\begin{itemize}
\item[($i$)]
Choose --~out of the class $\nu$~-- the configuration with minimal gauge-action
as the representative $A^{(\nu)}_{\rm std}$.
\item[($ii$)]
Choose --~out of the class $\nu$~-- the configuration for which the
gauge-action is closest to the class-average $\bar S^{(\nu)}$ as the
representative $A^{(\nu)}_{\rm std}$.
\end{itemize}
\item
Use the fermion action and the method you consider both trustworthy and
efficient to compute the standard-determinants
$\det((\Dsl^{(\nu)}_{\rm std}\!+\!M)/(\,\pas\!+\!M))$.
\item
For each of the higher topological sectors either include the corresponding
determinant computed in step 3 into the measurement or eliminate the
corresponding fraction of configurations from that sector.
\end{enumerate}
A few comments shall be added:

In step 1 the choice which method is used for determining $\nu$ --~though not
being part of the very definition of ``topologically unquenched QCD''~-- has
a crucial influence on whether the proposed method may be competitive or not.
We shall collect a few comments about this point in the appendix.

In step 2 the decision which criterion --~($i$) or ($ii$)~-- is used in order
to select the reference configurations may or may not have a big influence on
the outcome of the ``topologically unquenched'' measurement.
We emphasize that for either choice there is a sound theoretical motivation.
Strategy ($i$) --~choose the configuration which minimizes the gauge-part of
the total action~-- is nothing but the semiclassical ansatz being pushed to
account for topology:
Within each sector, the determinant is exact for the configuration having
least gauge-action, i.e. for the one which, in a semiclassical treatment,
gives the dominant contribution of that sector to the path-integral.
Strategy ($ii$) --~choose the configuration which in its gauge-action is
closest to the class-average of the sector~-- takes into account that in a
Monte Carlo simulation the system as a whole doesn't try to minimize the total
action-density but rather the free-energy-density:
The configuration which is most typical in a certain sector is not the one
with minimal action but the one which has additional instanton-antiinstanton
pairs and topologically trivial excitations such as to find an optimum between
the additional amount of action to be paid and the additional amount of entropy
to be gained.
It is the very aim of the second strategy to choose a background
which realizes such an optimum pay-off as reference-configuration.
A discussion in which regime of quark-masses and box-lengths the two strategies
are expected to lead to comparable results and in which regime only one of
them may be trustworthy follows near the end of the article.
Note finally that in the present form either strategy for selecting the
reference-backgrounds tries to find, in each sector, a configuration which
minimizes the action or maximizes the likelyhood to occur with respect to
the gluons only, none of them takes into account the shift --~both in action
and entropy~-- which is brought in by the fermions.

In step 3 one has to perform an ab-initio computation of the determinant
$\det((\Dsl^{(\nu)}_{\rm std}+M)/(\,\pas\!+\!M))$ w.r.t. two
backgrounds which cannot be continuously deformed into each other and which,
for this reason, are not close to each other.
This means that any approximate evaluation of the determinant which is
tantamount to an expansion in $\de A$ is necessarily inadequate.
Nevertheless, the task can be achieved -- e.g. by the eigenvalue method.

In step 4 the previously quenched sample gets modified so as to look more like
a full-QCD sample -- at least w.r.t. the distribution of winding numbers.
For QCD in a finite box Leutwyler and Smilga have shown that in the regime
\footnote{$\Sigma\!=\!\lim_{m\rightarrow0}\lim_{V\rightarrow\infty}
|\<\psb\ps\>|$, where $m_i\!=\!m\;\forall i$ for simplicity; note that
$V\Sigma m\rightarrow\infty$ when $m\rightarrow 0$ as the box has to
be scaled accordingly: $L\simeq 1/M_\pi, M_\pi^2\simeq m \Lambda_{\rm had}$}
$V\Sigma m\gg1$ the distribution of topological indices is gaussian with width
\cite{LeutwylerSmilga}
\beq
\<\nu^2\>=V\Sigma m/N_f
\label{topwidth}
\eeq
where $m/N_f$ is to be replaced by the reduced mass for unequally massive
flavors. In other words: By its width, this distribution knows about the
size~\footnote{The numerical value for $\Sigma$ is of course scheme- and
scale-dependent.} of the chiral condensate and thus about the amount of
dynamical symmetry breaking in QCD.
In quenched QCD the distribution is much broader as there is no determinant to
suppress the higher sectors and, as a consequence, there  is ``too much''
chiral symmetry breaking (which coincides with the observation that there is
one pseudo-Goldstone boson in excess \cite{QuenchedQCD}).
The ``topologically unquenched'' approximation introduces for each of the
higher sectors an average-determinant which reduces its weight.
As a result, the total distribution of topological indices gets narrower than
(strategy ($i$)) or as wide as (strategy($ii$)) the corresponding full-QCD
distribution (for details see below).
Whether the distribution being too narrow is equally unwelcome as being too
wide or whether the limited amount of dynamical symmetry breaking can be
compensated by somewhat larger quark-masses is an interesting question to be
studied numerically.

There are two aspects in which the algorithm above should be improved.
First, as the reweighting procedure results in eliminating a huge fraction
of the configurations from the previously quenched sample the ``topologically
unquenched'' determinant should get included into the measure right in the
beginning.
In addition, the more realistic strategy for choosing the reference
backgrounds should try to choose, in each sector, a configuration  which is
``most typical'' in the sense~\footnote{Anticipating that in a given sector
a histogram-plot of the configurations as a function~of
$S^{(\nu)}_{\rm gauge}$ (for a quenched sample) or of
$S^{(\nu)}_{\rm gauge}-\log(\det(\Dsl\!+\!M)/\det(\pas\!+\!M))$
(in case of full QCD) shows roughly a gaussian distribution (cf. op.3  in
\cite{LatticeIndex}), the configuration which differs in its gauge- or
total action from the corresponding class-average by the smallest absolute
amount is considered the ``most-typical'' configuration of that sector in the
sense of quenched or full QCD, respectively.}
of full QCD (strategy ($ii'$)) rather than in the sense of quenched QCD
(strategy ($ii$)).
These topics are addressed in due course.

The fact that the functional determinant in (\ref{for}) is a number which
depends only on the total topological charge of the background configuration
but not on its other details suggests that one could try to precompute the
determinants on artificially constructed backgrounds prior to running the
``topologically unquenched'' simulation.
From the Leutwyler-Smilga result (\ref{topwidth}) (which applies to full QCD)
one has an estimate of how many determinants will eventually be needed.
The backgrounds may be constructed following either of the two strategies
mentioned above.
Within the strictly semiclassical strategy --~ choice~($i$)~-- the
reference-backgrounds are gotten in a rather simple way:
Place $\nu$ copies of a sufficiently large
(i.e. $\rh\simeq0.3{\rm fm}>\rh_{\rm thr}$) single-instanton solution
(for $\nu\!>\!0$) on the lattice and cool this background (with a sufficiently
perfect action) in order to allow the instantons to adjust their positions and
their relative orientations in colour-space.
Within the more realistic strategy --~choice~($ii'$) which takes into account
the competing effects of increased action versus increased entropy~-- the
backgrounds are constructed as follows:
Place $\nu$ instantons (for $\nu\!>\!0$) with typical sizes (i.e.
$\rh\simeq0.3{\rm fm}$) randomly on the lattice plus additional instanton-
antiinstanton pairs such as to achieve an instanton density of $1{\rm fm}^{-4}$
\cite{SchaferShuryak, Negele}.
Optionally, dress this background with thermal fluctuations by
applying a reasonable number of heating-steps (monitoring $\nu$ in
order to guarantee that it stays unchanged).
Thus a pure Metropolis algorithm might look as follows:
\begin{enumerate}
\item
Determine from the box-length and the quark-masses at which your simulation
shall be done --~via (\ref{topwidth})~-- the $\nu$ needed
\footnote{The fact that a periodic torus doesn't support a selfdual solution
with $\nu\!=\!\pm1$ \cite{BraamBaal} is completely irrelevant in the context of
Lattice-QCD: When the box length is much larger than the instanton size a wide
window of meta-stability opens which, for all practical purposes, amounts to
stable single-(anti-) instanton solutions. Moreover, by using twisted boundary
conditions in at least one spatial direction the problem is completely avoided.}
and construct the corresponding $A^{(\nu)}_{\rm std}$ --~according to philosophy
$(i)$ or $(ii')$~-- as indicated above.
\item
Use the fermion action and method you trust to evaluate, on the backgrounds
constructed in step 1, the determinants
$\det((\Dsl^{(\nu)}_{\rm std}+M)/(\,\pas\!+\!M))$.
\item
Run the updating algorithm you trust; determine for each newly proposed
configuration its topological index (via the method you trust) and include
the ratio of the corresponding two standard determinants computed in step 2
into the Metropolis acceptance test, i.e. base this test on
\beq
\Delta S\equiv
S^{(\nu),{\rm new}}_{\rm glue}-
S^{(\nu),{\rm old}}_{\rm glue}-
\log(
{\det(\Dsl^{(\nu,{\rm new})}_{\rm std}+M)\ovr
\det(\Dsl^{(\nu,\,\,{\rm old}\,)}_{\rm std}+M)}
)
\label{topunquenchmetropolis}
\eeq
where the latter contribution is nonzero only if there is a change
$\nu_1\!\rightarrow\!\nu_2$.
\end{enumerate}
A few comments shall be added:

The alert reader will have realized that the prescription for constructing the
standard-backgrounds makes use of knowledge about the size-distribution and
partly about the density of (anti-) instantons in full QCD which was won in
previous lattice-studies.
In other words: The ``topologically unquenched'' simulation as outlined above
is not entirely from first principles.
In fact, the approximation depends --~strictly speaking~-- on the
strategy-intrinsic quality of the artificially constructed backgrounds which
got used for computing the determinants.
In particular, constructing the reference backgrounds according to strategy
($ii'$) is a non-trivial task, since it means that one has to make an a-priori
guess concerning the ``most typical'' configurations in full QCD.
Nevertheless, it should be stressed that the approximation is supposed to be 
fairly insensitive to the details of this choice -- provided the ``mistake''
is done uniformly in all sectors: Though strategies ($i$) and ($ii'$) end up
constructing reference-backgrounds which look quite different (very smooth in
the first case versus pretty rough in the second case) the final
``topologically unquenched'' sample may, in case the volume is not too large,
still be pretty much the same in either case -- the only thing which matters is
the (strategy-intrinsic) ratio of determinants computed in step 3 (for details
see below).

There is one more issue which shall be mentioned briefly:
In full QCD the HMC-algorithm was found to have serious problems in fluctuating
between the different topological sectors if quarks get sufficiently light
\cite{DecorrelationTime}.
It may well be that newly suggested configurations which would cross
into another sector are unlikely to get accepted not because of an increase
in gluon action but because of a decrease of the log of the functional
determinant they would bring.
If this is true the situation is likely to be better in ``topologically
unquenched QCD'' as in this approximation the newly proposed configuration
is judged not on the basis of its own determinant but of the
collective determinant of the new sector.
Needless to say that, in case a numerical investigation finds such an improved
behaviour, this would be interesting in the context of full QCD as well.


\section{Cost Estimate}

At this point it might be worth convincing the reader that the extra-costs
(as compared to a quenched run) brought by the ``topologically unquenched''
approximation are, in fact, rather moderate.

Regardless of the strategy chosen for selecting or constructing the
standard-backgrounds, the overhead as compared to a quenched simulation
results from the CPU-time spent on determining $\nu$ for every newly proposed
configuration and from the determinants which get evaluated.
There is, however, an important difference between these two cost-factors:
Preparing the reference-backgrounds and computing the determinants is a fixed
investment which is given by $L, a, m$ only -- it is independent of the length
of the simulation.
On the other hand, determining for each configuration its index $\nu$ gives
rise to costs which grow linearly in simulation-time and thus provide the main
overhead in a long run.

It is obvious that the method chosen for determining the topological index
will have the greatest impact on the overall-performance of ``topologically
unquenched QCD'' -- irrespective of whether strategy ($i$) or ($ii'$) is
implemented.
In the appendix we advocated a field-theoretical definition with little or no
cooling at all.
In the latter case at least half of the potentially suggestible backgrounds
have to be tossed away as they can't get an index assigned.
For the former variety we guess --~based on the evidence given in
\cite{Forcrand} how quickly cooling with an ``over-improved'' action tends to
pin down $g^2/(32\pi^2)\int G^a_{\mu\nu}\til G^a_{\mu\nu}\;dx$ near an integer
(say 5 sweeps to be within 2.99 and 3.01, etc.)~-- that applying $O(3)$ cooling
sweeps is sufficient to determine $\nu$.
Thus, in an approximation where a cooling-sweep is considered twice as
expensive as a complete update, the overhead from $\nu$-determinations is
roughly a factor 2...6 over a quenched simulation.

In addition, there are costs for computing the determinants.
In order to give an estimate, assume that a ``topologically unquenched''
simulation shall be done with up- and down-quarks having realistic masses.
Demanding that the pion ($M_\pi\simeq140{\rm MeV}$) fits three times into the
box, one ends up requiring $L\simeq4.4{\rm fm}$.
Trusting the usual values \footnote{Values for $\Sigma$ and $m$
are scheme- and scale-dependent, but the product is RG-invariant.}
$\Sigma\simeq(200{\rm MeV})^3$ and $m\simeq5{\rm MeV}$, one realizes that the
simulation will be in the regime $V\Sigma m\simeq10$.
From equ. (\ref{topwidth}) $\<\nu^2\>\simeq5$, i.e. the Leutwyler Smilga
analysis predicts $\#(\nu)\propto e^{-\nu^2/(2\cdot5)}$ \cite{LeutwylerSmilga}.
This means that the sectors $\nu=-10\ldots10$ will be sufficient for
simulations producing up to 10'000 independent configurations.
Hence $O(20)$ standard-determinants \footnote{This number could be cut by a
factor of 2 since the standard determinant is symmetric $\nu\leftrightarrow
-\nu$ but it may be advantageous to be able to check this explicitly.} must be
computed.
Strategy ($i$) is relatively cheap: The artificially constructed backgrounds
tend to be extremely smooth, thus the lowest $O(50)$ eigenvalues are expected
to give a reliable estimate for these determinants.
Strategy ($ii'$) somewhat more expensive: There are still $O(20)$ determinants
only which shall get evaluated, but this time the backgrounds do have
high-temperature fluctuations, thus the lowest $O(150)$ eigenvalues of the
Dirac operator will be required.

In summary, the ``topologically unquenched'' approximation is expected to
consume about one order of magnitude more CPU-time than a quenched run,
i.e. it might be considered an alternative to a high-statistics quenched study.


\section{Qualitative Aspects}

The proposal for ``topologically unquenched QCD'' comes with two strategies
--~($i$) and ($ii/ii'$)~-- for selecting or constructing the reference
backgrounds.
In order to judge the quality of these approximations we list the various kinds
of damage one may do to a configuration before computing ``its'' determinant
\begin{itemize}
\item
removing (topologically trivial) high-frequency excitations
\item
removing instantons and antiinstantons pairwise
\item
removing the remaining (anti-)instantons (of one kind)
\end{itemize}
and declare our firm belief that the harm done by these modifications
increases from top to bottom \footnote{Some supporting evidence can be found
in \cite{Negele} and references cited therein, but the aim of this proposal is
to persuade at least one collaboration to implement ``topologically unquenched
QCD'' in its various forms, as this would (hopefully) provide a direct
verification of the claim.}.
Accepting this point of view, it is easy to rank the approximations to full
QCD mentioned so far (worst first -- best last):
\begin{itemize}
\item[($\,o\,$)]
removing all sorts of excitations thoroughly -- the determinant gets
``evalu\-ated'' on an entirely flat background: this is the quenched
approximation.
\item[($\,i\,$)]
removing all topologically trivial excitations and as many
instanton-anti\-instanton pairs as possible -- the determinant gets evaluated on
an extremely smooth background which reflects nothing but the excess of
instantons over antiinstantons (or vice versa) in the original configuration:
this is ``topologically unquenched QCD'' with strategy $(i)$.
\item[($ii'$)]
removing or replacing the topologically trivial excitations and
accounting for the topologically nontrivial ones by an estimated number of
instanton-antiinstanton-pairs and the correct number of single-(anti-)
instantons: this is ``topologically unquenched QCD'' with strategy $(ii')$.
\end{itemize}
Knowing that the determinant depends in a particularly sensitive way on the
precise locations of the low-lying eigenvalues of the Dirac operator,
approximations ($o$) and ($i$) seem to amount to serious mutilations
of full QCD:
Since the work by Banks and Casher it is known that the phenomenon of
chiral symmetry breaking in QCD is associated with a coalescence of
low-lying eigenmodes of the Dirac operator near zero virtuality (see
\cite{LeutwylerSmilga} and references therein).
These low-lying modes are usually thought of as being the descendents of
the exact zero-modes (in the continuum) of the (anti-) instantons present
in the configuration.
Thus eliminating all or most of the (anti-) instantons seems to cause
serious harm.
We have two comments on this:
First, the density near zero virtuality is not directly proportional
to the (anti-)instanton density -- in fact, in the Instanton Liquid
Model the chiral symmetry restoration at high temperature is due to
instanton-antiinstanton pairs aligning in euclidean time direction
\cite{SchaferShuryak, InstantonLiquidModel}, i.e. the (anti-)instanton
density isn't altered that dramatically near the chiral phase-transition.
Second, we recall that the elimination of excitations concerns only the
``copy'' of the configuration used for the ``determinant-evaluation'' and not
the configuration itself -- otherwise it would be a mystery why the quenched
approximation ($o$) could have anything to do with full QCD.

We do not only expect strategy ($i$) to do better than the quenched
approximation ($o$), but we also expect strategy ($ii'$) to do better than
($i$) for the following reason:
The ``topologically unquenched'' determinant leads to a suppression of the
higher topological sectors, which --~in view of (\ref{topwidth})~-- is welcome,
but within strategy ($i$) this suppression is likely to be much too severe in
large boxes.
This prediction follows from the fact that the determinant introduced in
strategy ($i$) is exact for the background which, from the classical point
of view, dominates that sector.
The point is that this semiclassical treatment is indeed adequate for
sufficiently small coupling-constant, i.e. in a ridiculously small box.
In a larger volume the effective coupling strength increases and strategy ($i$)
is unable to account for this change.
To see this more clearly we stipulate the validity of the index theorem on the
lattice \cite{LatticeIndex} which allows us to rewrite the two determinants
appearing in~(\ref{thr}) using the Vafa-Witten representation \cite{VafaWitten}
\bea
{\det(\Dsl^{(\nu)}_{\rm std}\!+\!M)\ovr\det(\,\pas\!+\!M)}
&=&
\prod\limits_{i=1}^{N_{\rm f}}\;
m_i^{|\nu|}\!\cdot\!
{\prod_{\la>0}\det(\la^{(\nu)\,2}_{\rm std}+m_i^2)\ovr
\prod_{\la>0}\det(\la^{\,2}_{\rm free}+m_i^2)}\!\!
\label{vafawittentopo}
\\
\nonumber
\\
{\det(\Dsl^{(\nu)}\!+\!M)\ovr\det(\Dsl^{(\nu)}_{\rm std}\!+\!M)}
&=&
\prod\limits_{i=1}^{N_{\rm f}}\;
{\prod_{\la>0}\ \det(\la^{(\nu)\,2}+m_i^2)\ovr
\prod_{\la>0}\ \det(\la^{(\nu)\,2}_{\rm std}+m_i^2)}
\quad.
\label{vafawittencont}
\eea
Strategy ($i$) retains a determinant (\ref{vafawittentopo}) which is
appropriate in a small volume and thus strongly suppresses the higher
topological sectors.
As it comes to larger volumes, the semiclassical treatment breaks down
and the ``continuous'' determinant (\ref{vafawittencont}) reduces the
suppression of the higher sectors caused by the original form of
(\ref{vafawittentopo}) -- in full QCD, but not within strategy ($i$).
The virtue of strategy ($ii'$) is that this change is accounted for
by successively redefining the standard-backgrounds used in
(\ref{vafawittentopo}).
In other words: Within strategy ($ii'$) parts which would belong to
(\ref{vafawittencont}) in ($i$) are gradually reshuffled into the
``topological'' part (\ref{vafawittentopo}) as the box-volume increases.
As a consequence, strategy ($i$) might be trustworthy only as long as
$V\Sigma m\leq 1$; strategy ($ii'$), on the other hand, is hoped to give a
reasonable approximation to full QCD even in the regime $V\Sigma m\gg1$.

The Vafa-Witten-representation of the functional determinant is interesting in
yet another respect as it shows that the two factors (\ref{vafawittentopo}) and
(\ref{vafawittencont}) do have different structures.
It is the prefactor $m^{|\nu|}$ ($m_i\!=\!m\;\forall m_i$ for simplicity)
which makes the difference.
In QCD, this prefactor is known to cause the strong sup\-pression of nonzero
winding numbers in the limit $V\Sigma m\ll1$ \cite{LeutwylerSmilga}.
In other words: This prefactor accounts for the fact that chiral symmetry gets
restored if the chiral limit is performed in a finite volume.
The fact that this prefactor is still around in the ``topologically unquenched''
approximation (with either choice for the reference-backgrounds) means that
TU-QCD (unlike Q-QCD) gets the phenomenon of chiral symmetry restoration 
in a finite box qualitatively right.

Finally, the fact that the number of virtual quark-loops is not restricted in
TU-QCD means that there is an infinite number of diagrams contributing to the
$\et\pri$-propagator (not just the connected and the hairpin diagram as in
Q-QCD) and this propagator may even be well-defined in the field-theoretic
sense.


\section{Summary}

In this letter it is proposed to factorize the QCD fermion functional
determinant into two factors, the first one referring to a standard background
in the actual topological sector, the second one describing the effect of the 
deviation of the actual configuration from that reference background.
Then ``topologically unquenched QCD'' is defined to take the ``topological''
factor into account (with exactly the same quark-mass as in the propagator) and
to set the ``continuous'' factor to one.
In this sense the quarks happen to be fully dynamical (the number of
quark-loops not being restricted whatsoever) but to interact in a way which
pays no attention to the details of the gluon-configuration but just to its
index which, in turn, is sensitive to the topologically nontrivial excitations
only.

The proposal comes along with two strategies of how to select or construct the
reference-backgrounds on which the standard-determinants get evaluated:
One of them adopts a semiclassical point of view, the other one tries to
choose a configuration which is as likely to occur as possible.
We have given an estimate in which regimes of quark-masses and box-lengths
one or the both of these two strategies may render ``topologically unquenched
QCD'' an approximation which is reasonably close to full QCD and we have argued
that costs in terms of CPU-time are implied which are roughly one order
of magnitude higher than the costs of a quenched run.
Thus in ``topologically unquenched QCD'' direct simulations at physical
$M_\pi/M_\rh$ are possible on present day's machines.

The ``topologically unquenched'' approximation is expected to reduce
the problem of ``exceptional configurations'' encountered in quenched QCD but
it is unlikely to eliminate it completely:
In ``topologically unquenched QCD'' configurations get suppressed as compared
to quenched QCD, but the determinant which achieves this suppression doesn't
pay attention to anything but the number of lattice-descendents of zero-modes
of the Dirac operator on that configuration.
The main difference as compared to quenched QCD lies in the distribution of
topological indices: For sufficiently large volume strategy ($i$) effectively
acts as a constraint to the topologically trivial sector whereas strategy
($ii'$) is expected to give rise to a gaussian distribution with
full-QCD-appropriate width.
 
In spite of how attractive these theoretical aspects may look, we feel that an
honest judgement of how useful the ``topologically unquenched'' approximation
may be is likely to be possible only after it has been implemented.
Nevertheless, even for the case of full QCD it might prove useful to split the
functional determinant into a topologically trivial and a topologically
nontrivial factor and to evaluate the two of them by different techniques and
to different accuracies.


\section*{Appendix: Determining the Topological Indices}

Determining the topological index of a gluon background in a way which produces
results quickly and reliably is so crucial to the overall-performance of
``topologically unquenched QCD'' that the literature on this point shall be
briefly reviewed.
We are aware of four methods to determine the topological index
\cite{LatticeIndex}:
\begin{itemize}
\item[($a\,$)]
The local field-theoretic method: The index is defined through
\beq
\nu={g^2\ovr32\pi^2}\int G_{\mu\nu}^a\til G_{\mu\nu}^a\;d^4x
\label{metha}
\eeq
where $G_{\mu\nu}^a\til G_{\mu\nu}^a$ is implemented on the lattice by any
operator approaching this limit under $a\to 0$ (the simplest choice being
tantamount to adding up the sines of the plaquette-angles).
\item[($b\,$)]
The global field-theoretic method: The index is defined through
\beq
\nu={g^2\ovr32\pi^2}\int_S K_\mu d\si_\mu\;\;\mbox{with}\;\;
K_\mu=2\ep_{\mu\nu\si\rh}
(A_\nu^a \pad_\si A_\rh^a-{2\ovr3}g f^{abc} A_\nu^a A_\si^b A_\rh^c)
\label{methb}
\eeq
where the surface-integral is over $S^3$ at infinity and $K_\mu$ is implemented
by any operator having the appropriate continuum-limit.
\item[($c\,$)]
The index-theorem based method: The index is defined through
\beq
\nu=n_- - n_+
\label{methc}
\eeq
where $n_- (n_+)$ denotes the number of lattice-descendents of lefthanded
(righthanded) exact zero-modes of $\Dsl$ in the continuum.
\item[($d\,$)]
The anomaly-based method: The index is defined through
\beq
\nu=\lim_{m\to 0}\;m\int\psd\gaf\ps\;d^4x
\label{methd}
\eeq
where $\lim_{m\to 0}$ implies critical tuning (for a Wilson-type action).
\end{itemize}
While many of these methods were recently tested and found to give rise to
comparable results (when implemented with sufficient care) for the topological
susceptibility in QCD \cite{IndexComparison} we believe that in our case
--~nothing being known about the spectrum of $\Dsl$ on the background at hand~--
method ($a$) is likely to determine the topological indices in the quickest
possible way.
Nevertheless, the fact that on the lattice the operator involved in ($a$)
undergoes thermal renormalization provides a challenge:
Simply integrating the Chern density, i.e. computing $g^2/(32\pi^2)\int
G^a_{\mu\nu}\til G^a_{\mu\nu}\;dx$ gives a value which is, in general, not
close to an integer. In fact, a histogram-plot over many configurations tends
to reveal accumulations near regularly displaced, non-integer values, e.g. near
$0,\pm0.7,\pm1.4$ etc.
There are two options of how to deal with this situation:

The first, simplistic, approach is just to define a ``confidence interval''
--~e.g. $\pm0.2$~-- around each of the values $0,\pm0.7,\pm1.4,\ldots$ and to
assign the configurations lying within these bounds the indices $\nu=
0,\pm1,\pm2\ldots$ etc. The remaining configurations which didn't get an index
assigned are then simply tossed away.

The second, more sophisticated, approach is to make use of the fact that
cooling a configuration is able to remove the effect brought in by thermal
renormalization: Cooling a set of gluon-configurations results in the peaks
(in the histogram plot) being shifted closer to the corresponding integers
and the valleys between the peaks getting thinned out under each sweep.

The problem, however, is that these two methods do not necessarily agree in
their results: A configuration which has, without cooling, a measured value of
$g^2/(32\pi^2)\int G^a_{\mu\nu}\til G^a_{\mu\nu}\;dx$ so close to $0.7$
(in our example) as to justify a classification as of $\nu=1$ in the simplistic
approach, may easily be assigned, after a few cooling-sweeps, the trivial
index $\nu=0$ in the sophisticated approach.
While this phenomenon certainly is annoying in practice, it is nothing we have
to worry about in principle, as its origin can be understood on rather simple
grounds and the disagreement is expected to disappear, once the lattice-spacing
is sufficiently small:
The mismatch is caused by the fact that besides eliminating the high frequency
noise (which causes the multiplicative renormalization), cooling also affects
the instanton content of a configuration.
The problem is that cooling does not only favour annihilation of
instanton-antiinstanton pairs (which is completely harmless in the present
context) but also tends to influence the size-distribution of the remaining
(anti-)instantons -- which may have a dramatic influence on $\nu$-assignments
on lattices with nowadays typical $a$.
The point is that under repeated cooling with the naive (Wilson) action, a
single-instanton solution shrinks monotonically until it finally falls
through the grid.
In order to prevent the cooling algorithm at least from loosing large
instantons one has to modify the action w.r.t. which cooling is done in such
a way that all instantons with a radius $\rh$ above a certain threshold
$\rh_{\rm thr}$ (typically $\rh_{\rm thr}\simeq 2.3a$) tend to get blown up
(``over-improved actions'') or stay constant (``perfect action'') under
a sweep.
The price to pay, however, is that the small ones ($\rh<\rh_{\rm thr}$) get
compressed and finally pushed through the grid even more efficiently than
under cooling with an unimproved action \cite{Forcrand}.
Thus determining the topological index of a configuration with cooling
(using an improved action) yields results for the integrated Chern density
which are sharply peaked near integer values but the procedure is
sensitive to instantons with $\rh>2.3a$ only.
On the other hand, determining the topological index by the first ``simplistic''
approach (no cooling being involved) has an inferior signal-to-noise ratio
(about half of the configurations can't be assigned an index and have to be
tossed away) but the advantage is that the procedure is sensitive to all
instantons the lattice can support (i.e. those with $\rh>0.7a$).
Hence, any potential disagreement between the simplistic and the sophisticated
assignment is naturally explained as being due to instantons with sizes between
$0.7a<\rh<2.3a$ (approximatively).
The statement that such a disagreement will disappear once the lattice-spacing
is small enough takes its origin from the fact that in an $SU(3)$-type
gauge-theory (with a realistic value for the string tension) the distribution
of (anti-)instantons as a function of their radius is sharply peaked around
$\rho_0\simeq 0.3{\rm fm}$ with small sizes suppressed according to
$n\propto(\rho/\rho_0)^6$ \cite{SchaferShuryak}.
For example, on a lattice with $a=0.13{\rm fm}$ a mismatch may arize from
instantons having sizes between $\rh\simeq 0.1{\rm fm}$ and $0.3{\rm fm}$,
which is a considerable fraction of all instantons.
Once the lattice-spacing is as small as e.g. $a=0.03{\rm fm}$ the percentage
of configurations for which the two varieties don't agree is supposed to be
extremely small, since in this case either sensitivity-threshold
($\rh_{\rm thr}\simeq0.02{\rm fm}$ versus $\rh_{\rm thr}\simeq0.07{\rm fm}$)
lies in the $(\rho/\rho_0)^6$-type suppressed tail of the distribution.


\subsection*{Acknowledgments}
\vspace{-2pt}
It is a pleasure to acknowledge a series of most delightful conversations with
Steve Sharpe. In addition, I would like to thank Philippe de Forcrand for
teaching me about various cooling algorithms.
\newline
This work is supported by the Swiss National Science Foundation (SNF).


\clearpage

\vspace{-4pt}
\begin{thebibliography}{17}
\vspace{-4pt}
\bibitem{Hamber}
H.Hamber, G.Parisi, Phys.Rev.Lett. {\bf 47}, 1792 (1981);
E.Marinari, G.Parisi, C.Rebi, Phys.Rev.Lett. {\bf 47}, 1795 (1981);
D.Weingarten, Phys.Lett. {\bf 109B}, 57 (1982).
\vspace{-1pt}
\bibitem{SESAM}
N.Eicker et al. (SESAM), Phys.Lett. {\bf B407}, 290 (1997).
\vspace{-1pt}
\bibitem{Morel}
A.Morel, J.Phys.{\bf 48}, 111 (1987).
\vspace{-1pt}
\bibitem{QuenchedQCD}
C.Bernard, M.Golterman, Phys.Rev. {\bf D46} 853 (1992) and Nucl.Phys.
Proc.Supp. {\bf 26}, 360 (1992);
S.Sharpe, Phys.Rev. {\bf D46}, 3146 (1992) and Nucl.Phys. Proc.Supp. {\bf 30},
213 (1993).
\vspace{-1pt}
\bibitem{PartiallyQuenchedQCD}
C.Bernard,M.Golterman, Phys.Rev. {\bf D49}, 486 (1994);
S.Sharpe, Phys.Rev. {\bf D56}, 7052 (1997);
M.Golterman, K.Leung, Phys.Rev. {\bf D57} 5703 (1998).
\vspace{-1pt}
\bibitem{MQA}
W.Bardeen et al, Phys.Rev. {\bf D57}, 1633 (1998).
\vspace{-1pt}
\bibitem{Topology}
A.Belavin, A.Polyakov, A.Schwartz, Y.Tyupkin, Phys.Lett.{\bf 59B}, 85 (1975);
G.t'Hooft, Phys.Rev.Lett. {\bf 37}, 8 (1976) and
Phys.Rev. {\bf D14}, 3432 (1976);
C.Callan, R.Dashen, D.Gross, Phys.Lett. {\bf 63B}, 334 (1976);
R.Bott, Bull. Soc.Math.France {\bf 84}, 251 (1956).
\vspace{-1pt}
\bibitem{LeutwylerSmilga}
H.Leutwyler, A.Smilga, Phys.Rev. {\bf D46}, 5607 (1992); A.Smilga,
hep-th/9503049.
\vspace{-1pt}
\bibitem{LatticeIndex}
J.Smit, J.C.Vink, Nucl.Phys. {\bf B286}, 485 (1987);
R.Narayanan, P.Vranas, Nucl.Phys. {\bf B506}, 373 (1997);
C.R.Gattringer, I.Hip, C.B.Lang, Phys. Lett. {\bf B409}, 371 (1997) and
Nucl.Phys. {\bf B508}, 329 (1997);
C.R.Gattringer, I.Hip, hep-lat/9712015;
P.Hasenfratz,V.Laliena,F.Niedermayer, Phys.Lett. {\bf B427}, 125 (1998);
P.Hernandez, hep-lat/9801035.
\vspace{-1pt}
\bibitem{SchaferShuryak}
T.Sch\"{a}fer, E.V.Shuryak, Rev.Mod.Phys. {\bf 70}, 323 (1998).
\vspace{-1pt}
\bibitem{Negele}
J.W.Negele, hep-lat/9709129, hep-lat/9804017 and hep-lat/9810053.
\vspace{-1pt}
\bibitem{BraamBaal}
P.Braam, P.vanBaal, Comm.Math.Phys. {\bf 122}, 267 (1989).
\vspace{-1pt}
\bibitem{DecorrelationTime}
B.All\'es et al, Phys.Lett. {\bf B389}, 107 (1996).
\vspace{-1pt}
\bibitem{Forcrand}
P.deForcrand, M.Garcia P\'erez, I-O.Stamatescu, Nucl.Phys. {\bf B499}, 409
(1997) and Nucl.Phys. Proc. Supp. {\bf 63}, 549 (1998).
\vspace{-1pt}
\bibitem{InstantonLiquidModel}
D.I.Diakonov, V.Petrov, Phys.Lett. {\bf 147B}, 351 (1984) and
Nucl.Phys. {\bf B245}, 259 (1984);
E.Shuryak, J.Verbaarschot, Nucl.Phys. {\bf B341}, 1 (1990) and
{\bf B410}, 55 (1993);
T.Sch\"{a}fer, E.Shuryak, J.Verbaarschot, Nucl.Phys. {\bf B412}, 143 (1994);
T.Sch\"{a}fer, E.Shuryak, Phys.Rev. {\bf D54}, 1099 (1996).
\vspace{-1pt}
\bibitem{VafaWitten}
C.Vafa, E.Witten, Nucl.Phys. {\bf B234}, 173 (1984) and
Comm.Math.Phys. {\bf 95}, 257 (1984). 
\vspace{-1pt}
\bibitem{IndexComparison}
B.All\'es, M.D'Elia, A.DiGiacomo, R.Kirchner, hep-lat/9711026.
\end{thebibliography}
\end{document}